\documentclass[12pt,preprint]{aastex}
\usepackage{lscape}

\begin{document}

\title{CHEMICAL ABUNDANCE STUDY OF ONE RED GIANT STAR IN NGC~5694 :
A GLOBULAR CLUSTER WITH DWARF SPHEROIDALS'
CHEMICAL SIGNATURE?\footnote{This paper includes data gathered with the 6.5 meter
Magellan Telescopes located at Las Campanas Observatory, Chile.}}

\author{Jae-Woo Lee\altaffilmark{1},
Mercedes L\'opez-Morales\altaffilmark{2,3}, \&
Bruce W.\ Carney\altaffilmark{4}}

\altaffiltext{1}{Department of Astronomy and Space Science,
Astrophysical Research Center for the Structure and
Evolution of the Cosmos,
Sejong University, 98 Gunja-Dong, Gwangjin-Gu, Seoul, 143-747, Korea;
jaewoolee@sejong.ac.kr}
\altaffiltext{2}{Carnegie Institution of Washington,
Department of Terrestrial Magnetism,
5241 Broad Branch Road NW, Washington, DC 20015, USA; mercedes@dtm.ciw.edu}
\altaffiltext{3}{Carnegie Fellow}
\altaffiltext{4}{Department of Physics \& Astronomy, University of
North Carolina, CB\#3255, Chapel Hill, NC 27599-3255; bruce@unc.edu}

\begin{abstract}
We report the abundance analysis of one red giant branch star
in the metal-poor outer halo globular cluster NGC~5694.
We obtain [Fe/H] = $-$1.93, based on the ionized lines, and
our metallicity measurement is in good agreement with previous estimates.
We find that [Ca+Ti/2Fe] and [Cu/Fe] of NGC~5694 are 
about 0.3 -- 0.4 dex lower than other globular clusters with similar
metallicities, but similar to 
some LMC clusters and stars in some dwarf spheroidal galaxies. 
Differences persist, however, in the abundances of neutron capture elements.
The unique chemical abundance pattern and the large Galactocentric
distance (30 kpc) and radial velocity ($-138.6 \pm 1.0$ km sec$^{-1}$)
indicate that NGC~5694 had an extragalactic origin.
\end{abstract}


\section{INTRODUCTION}
The cold dark matter model for cosmology predicts
a hierarchical formation mechanism for galaxies,
with smaller units accreting to construct larger ones
(e.g.\ Navarro, Frenk, \& White 1995). Signs of merger
fragments have been identified kinematically, especially
the Sagittarius dwarf galaxy (Ibata et al.\ 1994), and
probably the Monoceros Ring (Yanny et al.\ 2003). 
More substructure is predicted by the models but confirmations 
have proven elusive.

With the advent of large aperture telescopes, ``chemical tagging"
becomes a powerful technique to probe past merger histories.
As Freeman \& Bland-Hawthorn (2002) discussed, stars born in galaxies whose
star formation histories differ from those that have created the
bulk of the Galaxy's stars may still be discernible in
unusual element-to-iron ratios. For example, Cohen (2004) has
found a compelling link between Palomar~12 and the Sagittarius
dwarf, confirming the dynamical association found earlier 
by Dinescu et al.\ (2000).
Venn et al.\ (2004) summarized the unusual abundance
patterns found in the Galaxy's dwarf spheroidal (dSph) galaxy neighbors, 
demonstrating that the bulk of the Galactic halo did not come from such 
surviving systems.
Unique chemical abundance patterns of globular clusters and halo field stars
may become a primary method to identify common star formation
origins and histories.

NGC~5694 ($\ell = 331.1$, $b = +30.4$) is a metal-poor outer
halo globular cluster lying far from the Sun (Harris 1975;
Ortolani \& Gratton 1990), as well as far from the Galactic center.
Harris (1996) cites a Galactocentric distance of 29 kpc, E($B-V$) = 0.09, 
and a large radial velocity, $v_{\rm rad}$ = $-$144.1 km sec$^{-1}$.
The large velocity and distance led Harris (1975) and
Harris \& Hesser (1976) to suggest that
NGC~5694 has a hyperbolic orbit and it is not bound to our Galaxy,
or that the Galaxy contains considerable additional mass beyond
the solar orbit than simple model potentials indicate.
We have used the analytical model of the Galactic gravitational
potential from Allen \& Santillan (1991) to estimate a lower
limit to its apogalacticon distance.
We first re-estimate the cluster's distance,
employing a mean horizontal branch $V$ magnitude
of 18.5 (Harris 1996) and the $M_{\rm V}$ vs.\ [Fe/H] relation of
Cacciari (2003), finding $R_{\rm GC}$ = 30 kpc.
Based only on the radial velocity,
the cluster travels over 100 kpc from the Galactic center. Any significant
tangential velocity will likely increase this value further.
Inspired by the outer halo nature of the
cluster, we have begun a high-resolution spectroscopic
study of one red giant branch star in the cluster.
Our results suggest that NGC~5694 has a very distinctive elemental
abundance pattern, similar in some respects to those of nearby
dwarf spheroidal galaxies.

\section{OBSERVATIONS, DATA REDUCTION AND ANALYSIS}

We selected as our program star I-62 from the $BV$ photometry
by Harris (1975; $V$ = 15.55; $B-V$ = 1.32).
The star has quality ``A" 2MASS $JK$ photometry (Cutri et al.\ 2000; 
$K$ = 12.21, $J-K$ = 0.80).
The radial velocity measurements for red giant stars in NGC~5694
by Geisler et al (1995) using medium-resolution spectra
at the \ion{Ca}{2} infrared triplet showed that I-62
is a probable member of NGC~5694.

Our observations were carried out with the Magellan Clay
Telescope using the Magellan Inamori Kyocera
Echelle spectrograph (MIKE; Bernstein et al.\ 2003) on 6 July 2005.
A 0.\arcsec35 slit was used providing a resolving power of $\approx$ 50,000
in the red with wavelength coverage from 4950~\AA~to 7250~\AA,
based on the full-with half maximum of the Th-Ar emission features.
Four 2400s exposure were taken with this setting.
We also obtained a spectrum of a fast rotating hot star to remove
telluric absorption features.
We used {\sc MIKE Redux}\footnote{http://web.mit.edu/$\sim$burles/www/MIKE/}
to extract the spectra, and which effectively correct for the tilted slit.

Equivalent widths were measured mainly by direct integration of
each line profile using the SPLOT task in IRAF ECHELLE package.
We estimate our measurement error in equivalent width to be
$\pm2$ to $\pm4$ m\AA\ from the size of  noise features
in the spectra and our ability to determine the proper continuum level.

For our line selection, laboratory oscillator strengths
were adopted whenever possible,
with supplemental solar oscillator strength values.
We adopted the ``Uns\"old approximation" to account for
van der Waals line broadening with no enhancement
(Lee \& Carney 2002; Lee, Carney, \& Habgood 2005).
We included the effects of hyperfine splitting
for Mn, and both hyperfine and isotopic splitting for Cu and Ba.
We neglected such corrections for La and Eu
because the lines are very weak and the derived abundances are
therefore insensitive to damping.

The initial temperature of the program star was estimated
using its available $BVK$ photometry and the empirical color-temperature and 
bolometric correction-color relations given by 
Alonso, Arribas, \& Martinez-Roger (1999).
To estimate the star's relative to the Sun gravity using photometric data,
we used $\log g_{\sun}$ = 4.44 in cgs units, $M_{\rm bol,\sun}$ =
4.74 mag, and $T_{\rm eff,\sun}$ = 5777~K. Using the estimated cluster
distance and a stellar mass of $0.8 M_{\odot}$, we found $T_{\rm eff}$ = 4135~K
and log~$g$ = 0.6.

The abundance analysis was performed using the current version
of the local thermodynamic equilibrium (LTE) line analysis
program MOOG (Sneden 1973).
For input model atmospheres, we interpolated Kurucz models
using a program kindly supplied by A.\ McWilliam (2005, private communication).
Adopting the photometric temperature and surface gravity
as our initial values, we began by restricting the analysis to
those \ion{Fe}{1} lines with $\log$(W$_{\lambda}$/$\lambda$) $\leq$ $-5.2$
(i.e., for  the linear part of the curve of growth),
and comparing the abundances as a function of excitation potential.
New model atmospheres were computed with a slightly different effective
temperature until the slope of the $\log$~n(\ion{Fe}{1}) 
versus excitation potential relation was zero to within the uncertainties.
The stronger \ion{Fe}{1} lines were then added and the microturbulent
velocity $v_{\rm turb}$ altered until the $\log$ n(\ion{Fe}{1}) versus
$\log$(W$_{\lambda}$/$\lambda$) relation had no discernible slope.
We obtained $T_{\rm eff}$ = 4200~K and $v_{\rm turb}$ = 2.2 km sec$^{-1}$. 
[Fe/H] was found to be $-2.08 \pm 0.11$, based on the neutral iron lines, 
and $-1.93 \pm 0.07$, based on the ionized lines. 
Since metal-poor stars have much weaker metal-absorption in the ultraviolet (UV),
more non-local UV flux can penetrate from the deeper layers,
which leads to over-ionization of neutral lines.
Therefore Fe abundance derived from \ion{Fe}{1} lines for metal-poor stars
will always be underestimated, 
while Fe abundance derived from \ion{Fe}{2} lines remains unaffected
(Th\'evenin \& Idiart 1999; Ivans et al.\ 2001).
Our [Fe/H] values compare well with those
estimated by other means by Zinn \& West (1985; $-1.92$) and its
recalibration by Kraft \& Ivans (2003), based on reliance on only
the ionized lines for the calibrating clusters. 
Their derived [Fe/H] value, obtained using Kurucz models 
with convective overshoot turned on, as we have employed, was $-2.04$.

\section{RESULTS AND DISCUSSION}

\subsection{Radial Velocity}

We measured the heliocentric radial velocity of the program star
with respect to that of HD116713 using the IRAF FXCOR task and
obtained $-$138.6 $\pm$ 1.0 km sec$^{-1}$.
Our result is in good agreement with that of Geisler et al.\ (1995).
Neglecting the three most deviant velocities, the remaining ten stars in Geisler's sample
have a mean radial velocity of $-140.7 \pm 2.4$ km sec$^{-1}$ (the
error is that of the mean).
Our radial velocity measurement re-confirms that star I-62 is a member
of NGC~5694.

\subsection{Elemental Abundances}
Table~\ref{tab1} summarizes the elemental abundances of
NGC~5694 I-62 using the photometric surface gravity and 
the spectroscopic temperature.
The [el/Fe] ratios for neutral elements are estimated from [el/H]
and [\ion{Fe}{1}/H] ratios.
The [el/Fe] for singly ionized elements (\ion{Ti}{2}, \ion{Y}{2},
\ion{Ba}{2}, \ion{La}{2}, and \ion{Eu}{2})
are estimated from [el/H] and [\ion{Fe}{2}/H] ratios. 
This procedure follows the study of M5 giants by Ivans et al.\ (2001), 
and has been employed in our prior work as well 
(Lee \& Carney 2002; Lee, Carney, \& Habgood 2005). 
See also Johnson et al.\ (2006) for a discussion of the challenges
presented in comparing photometric and spectroscopic temperatures
and gravities.
In the Table, $n$ is the number of lines used for the calculations of
mean elemental abundances and $\sigma$ is the standard deviation per line.
Systematic errors, such as in adopted $gf$ values as a function
of excitation potential, which could lead to systematically
erroneous temperature estimates, are not included.
A more detailed discussion of elemental abundances will be presented elsewhere
in the future.

\subsection{Comparisons with other Stellar Systems}

Our results are based on the analysis of only one star, 
and the comparisons given below must be considered suggestive rather
than definitive. But NGC~5694 appears to be unusual, almost unique, 
in its chemical abundance pattern and warrants further study.

NGC~5694 I-62 is deficient in $\alpha$-elements, in particular Ca and Ti, and
the iron-peak element, Cu, compared with other globular clusters in our Galaxy.
For [Ti/Fe], we adopt the unweighted average of
[\ion{Ti}{1}/\ion{Fe}{1}] and [\ion{Ti}{2}/\ion{Fe}{2}].
Use of neutral titanium lines may suffer from NLTE effects,
such as an over-ionization.
However, the results from \ion{Ti}{2} lines
also yield lower titanium abundance scales in our program star,
indicating that it is truly titanium deficient.
In Figure~1, we show  [Ca+Ti/2Fe] and [Cu/Fe] ratios as functions of [Fe/H].
We also show those of other globular clusters
in our Galaxy (Pritzl et al.\ 2005; Simmerer et al.\ 2003),
Large Magellanic Cloud (LMC) globular clusters (Johnson et al.\ 2006)
and nearby dSph galaxies (Shetrone et al.\ 2003).
The [Ca+Ti/2Fe] ratio of NGC~5694 I-62 is very similar to those of
Palomar~12 and Terzan~7, which are associated with the Sagittarius
dwarf galaxy (see Dinescu et al.\ 2000 regarding Palomar~12's association),
and Ruprecht~106, which has been suggested to have been
associated with the Magellanic Clouds (Lin \& Richer 1992).
On the other hand, other $\alpha$-elemental abundances, [Mg/Fe] and [Si/Fe],
appears to be normal.
The LMC cluster results from Johnson et al.\ (2005) also showed
[Mg/Fe] and [Si/Fe] ratios similar to those of globular clusters in our Galaxy.
Some iron-peak elemental abundances for NGC~5694 I-62,
[Mn/Fe] and [Ni/Fe], appear to be normal
(Sobeck et al.\ 2006; Gratton et al.\ 2004) relative to other clusters.
However, the [Cu/Fe] ratio of NGC~5694 I-62 is $\approx$ 0.4 dex
lower than those of globular clusters studied by Simmerer et al.\ (2003)
at [Fe/H] $\approx$ $-$2.0 dex
and the nearby dSphs studied by Shetrone et al.\ (2003).
Some RGB stars in the Sculptor dSph galaxy\footnote{
Note that Sculptor dSph is the oldest dSph galaxy studied by
Shetrone et al.\ (2003).  However, it appears to have several Gyr of
active star formation and its star formation ended about 4 Gyr ago.
On the other hand other dSphs have had even more extended periods
of star formation, stopping only in the last 1 or 2 Gyr
(Dolphin 2002; Tolstoy et al.\ 2003).
As noted by the referee, proto-galactic fragments that disrupted very early
in the first few Gyr would resemble only the oldest stellar populations
in dSphs, which is not the dominant population in the dSphs
because of their slow, but steady, star formation rates.}
and LMC appear to have similar [Cu/Fe] ratios.

The $\alpha$-elements are predominantly synthesized during
the SNe~II shell-burning at the end of the lives of massive stars.
Most Cu appears to be synthesized by an $s$ process in massive stars.
(The relative importance of SN~Ia for Cu abundance remains uncertain,
according to Clayton 2003.)
Since NGC~5694 has a very old age (De Angeli et al.\ 2005),
the contributions from SNe~Ia are unlikely to be significant,
since such events are not thought to appear until $10^9$ or
more years following the beginning of star formation.
Further, the low [Cu/Fe] ratio of NGC~5694 cannot be understood by
a metallicity-dependent yield from SNe~Ia, which appears to be more important
in more metal-rich regimes (e.g.\ McWilliam \& Smecker-Hane 2005).
One possible explanation would be that NGC~5694 formed from a proto-globular
cluster cloud which was contaminated by relatively rare, massive SNe~II
(e.g.\ Tolstoy et al.\ 2003).
This suggests that NGC~5694 formed in very different environment
than the bulk of globular clusters in our Galaxy.

The neutron capture elements reveal even greater complexity. Venn
et al.\ (2004) and Johnson et al.\ (2006) have noted the low
abundances of [$\alpha$/Fe] and [Cu/Fe] for dSphs and
LMC clusters compared to Galactic halo field and
globular cluster stars, and Venn et al.\ (2004) drew special
attention to [Ba/Y] as a significant difference as well. The
LMC clusters studied by Johnson et al.\ (2006)
do not share this trend, having solar [Y/Fe] ratios, as found
in the Galactic halo, but somewhat enhanced [Ba/Fe]. NGC~5694 I-62
has a very low [Y/Fe] value, like the dwarf spheroidal galaxies, but
its [Ba/Fe] ratio is lower, and lower than the LMC clusters as well.
[Eu/Fe] is only slightly super-solar, resulting in a [Ba/Eu] ratio
well below the LMC clusters or the dSphs.

In short, NGC~5694 is similar in some chemical abundance ratios to
the LMC clusters and the dSphs, but a closer look
at the neutron capture elements suggests significant differences.

Is NGC~5694 related to any of the existing dSphs? Kinematically
the answer appears to be ``no". Lynden-Bell \& Lynden-Bell (1994) introduced
the concept of alignments of orbital poles, but did not identify NGC~5694
as related to any of the dwarf spheroidal galaxies. Majewski (1994)
employed radial velocities in addition to positional data and drew
a similar conclusion. Finally, space velocities have been employed
by Palma et al.\ (2002) to compare the clusters' motions, but,
unfortunately, NGC~5694 still does not have a measured proper
motion. We await such a measurement with keen interest, given that the
estimated large apogalacticon distance and unique chemical abundances
suggest that NGC~5694 formed independently of the bulk of the Galaxy,
and was captured subsequently. 

\acknowledgements
The authors thank David Yong and the anonymous referee 
for helpful discussions.
Support for this work was provided by the Korea Science
and Engineering Foundation (KOSEF) to the Astrophysical Research Center
for the Structure and Evolution of the Cosmos (ARCSEC), 
the Carnegie Institution of Washington through a Carnegie Fellowship
and a National Science Foundation grant AST-0305431 to the University
of North Carolina.


\clearpage

\begin{deluxetable}{lrlrrr}
\tablecaption{Elemental abundances.\label{tab1}}
\tablenum{1}
\tablewidth{0pc} \tablehead{ \multicolumn{1}{c}{Elem.} &
\multicolumn{1}{c}{Sun} & \multicolumn{1}{c}{} &
\multicolumn{1}{c}{} & \multicolumn{1}{c}{n} &
\multicolumn{1}{c}{$\sigma$} } \startdata
Fe & 7.52 & [Fe/H]$_{\rm{I}}$   & $-$2.08 & 40 & 0.11 \\

   &      & [Fe/H]$_{\rm{II}}$  & $-$1.93 &  7 & 0.07 \\

Mg & 7.58 & [Mg/Fe]             &   +0.15 &  2 & 0.03 \\

Si & 7.55 & [Si/Fe]             &   +0.32 &  2 & 0.01 \\

Ca & 6.36 & [Ca/Fe]             &   +0.00 &  7 & 0.06 \\

Ti & 4.99 & [Ti/Fe]$_{\rm{I}}$  &   +0.00 & 10 & 0.11 \\

   &      & [Ti/Fe]$_{\rm{II}}$ & $-$0.15 &  2 & 0.14 \\

   &    & [Ti/Fe]$_{\rm{mean}}$ & $-$0.07 & \nodata & \nodata \\

Mn & 5.39 & [Mn/Fe]             & $-$0.38 &  2 & 0.09 \\

Ni & 6.52 & [Ni/Fe]             & $-$0.10 &  3 & 0.09 \\

Cu & 4.21 & [Cu/Fe]             & $-$1.05 &  1 & \nodata \\

 Y & 2.24 & [Y/Fe]$_{\rm{II}}$  &  $-$0.58 &  2 & 0.02 \\

Ba & 2.13 & [Ba/Fe]$_{\rm{II}}$ & $-$0.52 & 3 & 0.09 \\

La & 1.22 & [La/Fe]$_{\rm{II}}$ &  $-$0.26 & 1 & \nodata \\

Eu & 0.51 & [Eu/Fe]$_{\rm{II}}$ &  +0.18 & 2 & 0.11 \\

\enddata
\end{deluxetable}


\clearpage

\begin{figure}
\figurenum{1}
\plotone{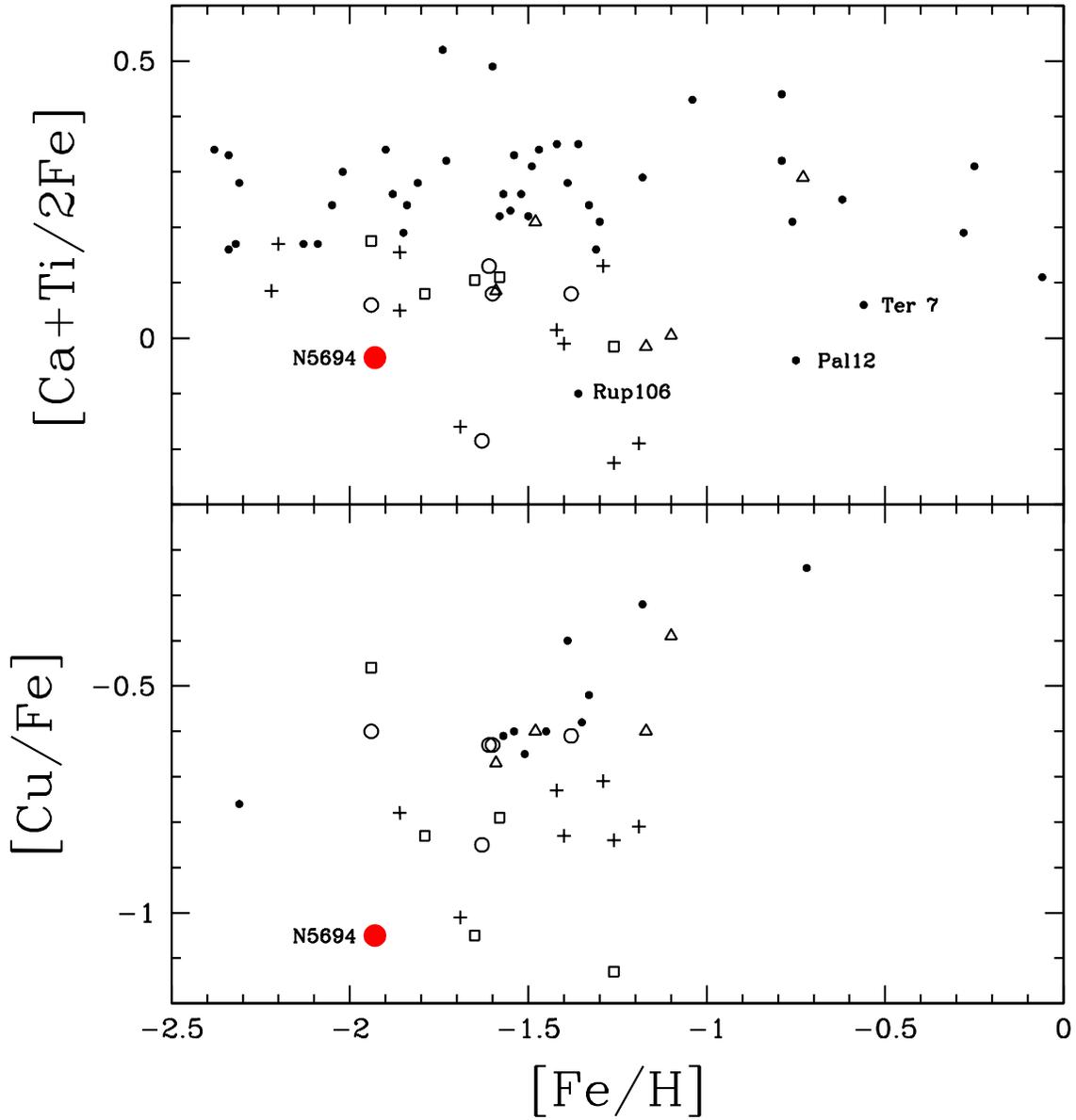}
\caption{[Ca+Ti/2Fe] and [Cu/Fe] ratios as functions of [Fe/H].
Dots represent globular clusters in our Galaxy, open circles the Carina dSph,
open triangles the Fornax dSph, open squares the Sculptor dSph,
and crosses LMC clusters.}\label{fig1}
\end{figure}

\begin{figure}
\figurenum{2}
\plotone{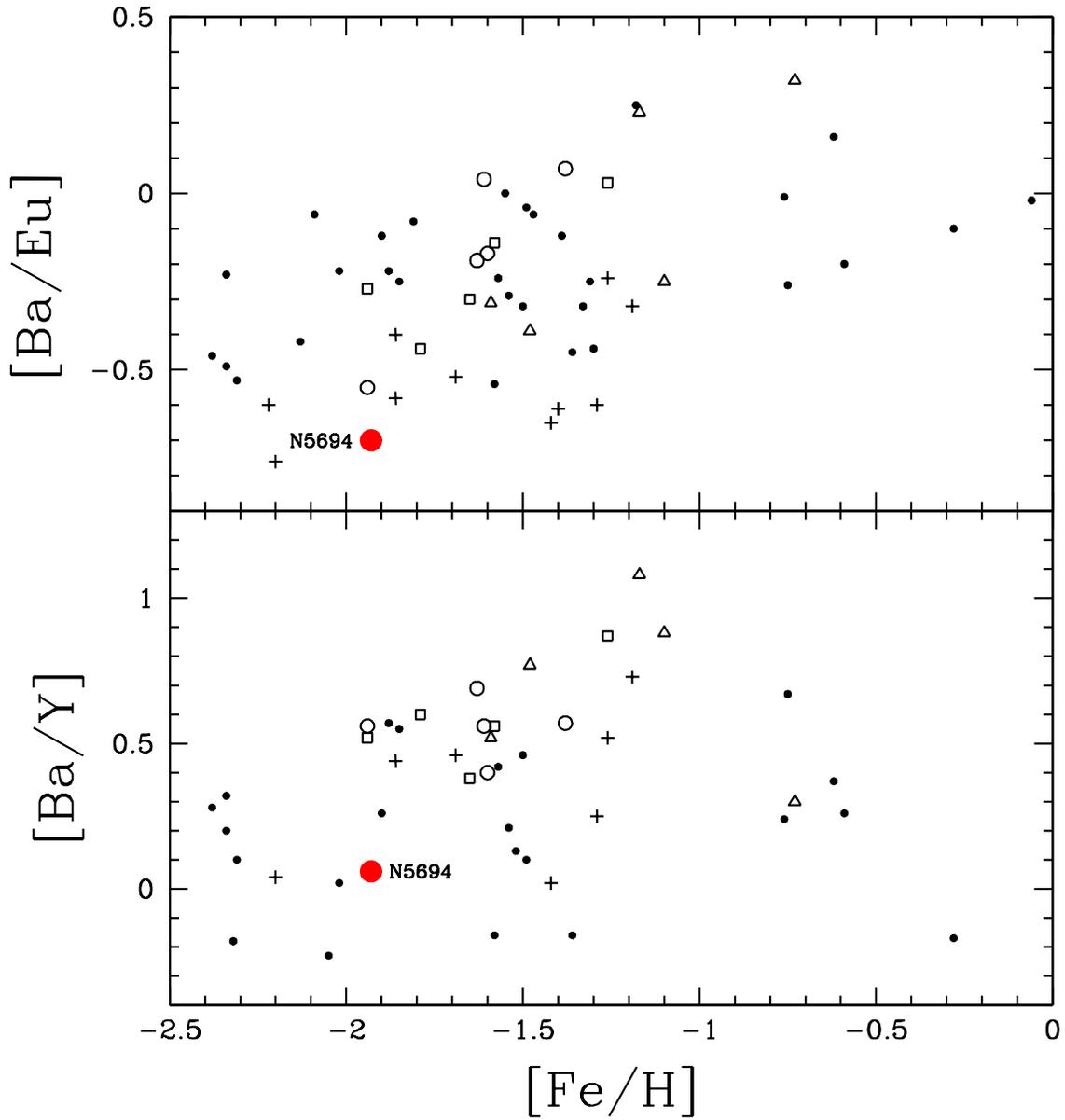}
\caption{[Ba/Eu] and [Ba/Y] ratios as functions of [Fe/H].
We use the same symbols as in Figure~\ref{fig1}.} \label{fig2}
\end{figure}

\end{document}